\documentclass[11pt]{article}
\usepackage{graphics}
\usepackage{rotating}
\usepackage{sao1}

\def\teff{$T_{\rm eff}$}
\def\lgg{$\log g$}

\def\vs{$v_{\rm e}\sin i$}

\newcommand{\bs}{$\langle B_{\rm s}\rangle$}
\newcommand{\kms}{km\,s$^{-1}$}

\newcommand{\ei}{E$_{\rm i}$}
\newcommand{\figps}[1]{\resizebox{\hsize}{!}{\rotatebox{0}{\includegraphics{#1}}}}

\newcommand{\fifps}[2]{\centering\resizebox{#1}{!}{\includegraphics{#2}}}
\newcommand{\firrps}[2]{\resizebox{#1}{!}{\rotatebox{-90}{\includegraphics{#2}}}}

\def\i{\,{\sc i}} \def\ii{\,{\sc ii}}

\begin{document}

\title{Ca isotopic anomaly in the atmospheres of Ap stars}
\author{T. Ryabchikova\inst{1,2} \and  O. Kochukhov\inst{3}
	\and S. Bagnulo\inst{4}}
\institute{Institute of Astronomy, Russian Academy of Sciences, Pyatnitskaya 48, 119017 Moscow, Russia\thanks{Based on 
observations collected at the European Southern Observatory, Paranal, Chile (ESO programme No. 68.D-0254)}
	    \and Department of Astronomy, University of Vienna, T\"urkenschanzstrasse 17, A-1180 Wien, Austria
	    \and Department of Astronomy and Space Physics, Uppsala University Box 515, SE-751 20 Uppsala, Sweden
	    \and European Southern Observatory, Casilla 19001, Santiago 19, Chile}
\maketitle 

\begin{abstract}
We present results of the Ca stratification analysis in the atmospheres of 21 magnetic chemically peculiar (Ap) stars. This analysis was based on the
spectral observations carried out with the UVES spectrograph attached to the 8-m VLT telescope. Ca was found to be strongly stratified in all stars
with different effective temperatures and magnetic field strengths. This element is overabundant by 1--1.5 dex below $\log\tau_{5000}\approx-1$ and
strongly depleted above $\log\tau_{5000}=-1.5$. Based on the overall Ca abundance distributions, we modelled a profile of the IR-triplet
Ca\ii\ $\lambda$~8498 line. It shows a significant contribution of the heavy isotopes $^{46}$Ca and $^{48}$Ca, which represent less than 1\% of the solar Ca
isotopic mixture. In Ap stars with the relatively small surface magnetic fields ($\le 4-5$ kG) the light $^{40}$Ca isotope is concentrated
close to the photosphere, while the heavy isotopes are pushed towards the outer layers. Isotopic separation disappears in the
atmospheres of stars with magnetic fields above 6--7 kG. The observed overall Ca stratification and isotopic anomalies may be
explained by a combined action of the radiatively-driven diffusion and the light-induced drift.  
\end{abstract}

\section{Introduction}
\label{intro}
 
After the pioneering work by Michaud (\cite{Michaud70}) particle diffusion in stellar envelopes and atmospheres
is considered as the main process responsible for the atmospheric abundance anomalies
in the peculiar stars of the Upper Main Sequence. Detailed diffusion calculations performed for a set
of chemical elements in the atmospheres of magnetic peculiar stars predicted an existence of abundance stratification.
For a small number of elements, including Ca, an effect of the stratified element distribution on the spectral line profiles 
was demonstrated in early studies (Borsenberger at al. \cite{BPM81}), but the absence
of high-resolution, high signal-to-noise spectroscopic observations did not allow the direct
comparison between the observations and diffusion calculations. This step was carried out by Babel (\cite{Babel92}), who calculated the
Ca abundance distribution in the atmosphere of magnetic star 53 Cam and showed that the unusual shape of Ca\ii\ K line
-- a sharp transition between the wide wings and extremely narrow core -- is a result of a step-like Ca distribution with abundance
decrease at $\log\tau_{5000}\approx-1$. Following Babel, the step-function approximation of the abundance distribution was commonly employed
in many stratification studies based on the observed profiles of spectral lines (Wade et al. \cite{WLRK03};
Ryabchikova et al. \cite{RPK02};  Ryabchikova et al. \cite{RLK05}; Ryabchikova et al. \cite{RRKB06}).

Ca was found to be stratified the same way as in 53 Cam (enhanced concentration of Ca below $\log\tau_{5000}\approx-1$ and
its depletion above this level) in all stars for which stratification analysis have been performed: $\beta$~CrB (Wade et al.
\cite{WLRK03}), $\gamma$~Equ (Ryabchikova et al. \cite{RPK02}), HD~204411 (Ryabchikova et al. \cite{RLK05}), HD~133792 
(Kochukhov et al. \cite{KTRM06}) and HD~144897 (Ryabchikova et al. \cite{RRKB06}). Recently another Ca anomaly was detected, first in the spectra of 
HgMn stars by Castelli \& Hubrig (\cite{CastH04})
and then in Ap stars by Cowley \& Hubrig (\cite{CH05} - CH). These authors found a displacement of the lines of Ca\ii\ IR triplet due to 
significant contribution of the heavy Ca isotopes. CH merely noted the fact of the presence of heavy isotopes, but they did not perform any quantitative 
analysis. This was done by Ryabchikova, Kochukhov \& Bagnulo, and the preliminary results were published in a review paper by Ryabchikova
(\cite{MONS05}). It was shown that the contribution of Ca heavy isotopes decreases with the increase of magnetic field strengths, and disappears
when the field exceeds 3 kG.     

In present study we give a detailed analysis of the Ca stratification in the atmospheres of magnetic
Ap stars of different temperatures and magnetic field strengths with the application to a modelling of the IR triplet 
Ca\ii\ $\lambda$~8498 line.

\section{Observations and data reduction.} 
\label{obs}

Twenty-one slowly rotating Ap stars were chosen for the Ca stratification analysis. For all but two stars, HD~24712 and HD~66318, 
high-resolution, high signal-to-noise-ratio spectra were obtained with the UVES instrument at the ESO VLT in the context of program 68.D-0254. The UVES instrument is
described by Dekker et al.\ (2000). The observations were carried out
using both available dichroic modes. In both the blue
arm and the red arm the slit width was set to 0.5$^{\prime\prime}$, for a spectral
resolution of about 80\,000.  The slit was oriented along the
parallactic angle, in order to minimize losses due to atmospheric
dispersion.  Almost the full wavelength interval from 3030 to 10400 \AA\
was observed except for a few gaps, the largest of which was at
5760-5835 \AA\ and 8550-8650 \AA.  In addition, there are several small
gaps, about 1\,nm each, due to the lack of overlapping between the
\'{e}chelle orders in the 860U setting. Spectra of HD~24712, HD~66318 and HD~61421 (Procyon) were obtained with the same setting 
and were extracted from ESO archive. Due to the gaps in spectral coverage, only one line of the Ca\ii\ IR triplet, $\lambda$~8498.023 \AA, could be observed in this UVES setting and is accessible for modelling. 

The Ca IR triplet line studied overlaps with the hydrogen lines from the Paschen series. 
Due to the difficulties of continuum normalization at the edge of observed spectral region, we have employed theoretical spectrum synthesis to establish the correct continuum level. In this procedure observations around Ca\ii\ $\lambda$~8498.023 \AA\ were adjusted, so that the pseudo-continuum of the Paschen line wings matches predictions of the theoretical spectrum synthesis.
 
The list of the program stars is given in Table~\ref{tbl1}. In addition, two stars, HD~27411 (A3m) and 
Procyon (HD~61421), were used as standards for the Ca isotopic study.

\begin{table*}[!th]
\caption{Fundamental parameters of target stars.\label{tbl1}}
\begin{tabular}{rlcccl}
\hline
HD~~~  & \teff     & \lgg & \vs~~     & \bs &Reference  \\
number & (K)	   &	  & (\kms)    & (kG)&		\\
\hline
\multicolumn{6}{c}{Program stars} \\
\hline
217522 & 6750& 4.30 & ~2.5   &$\le$1.5&Gelbmann (\cite{G98}) \\
122970 & 6930& 4.10 & ~5.5   & 2.5 &Ryabchikova et al. (\cite{RSH00}) \\  
 24712 & 7250& 4.30 & ~5.6   & 2.3 &Ryabchikova et al. (\cite{RLG97}) \\  
134214 & 7315& 4.45 & ~2.0   & 3.1 &this paper \\			 
   965 & 7500& 4.00 & ~3.0   & 4.4 &this paper \\
203932 & 7550& 4.34 & ~5.3   &$\le$1& Gelbmann et al \cite{GKWM97} \\
137949 & 7550& 4.30 & ~1.0   & 5.0 &Ryabchikova et al. (\cite{RNW04}) \\  
176232 & 7650& 4.00 & ~2.0   & 1.5 &Ryabchikova et al. (\cite{RSH00}) \\	       
 75445 & 7650& 4.00 & ~3.0   & 3.0 &Ryabchikova et al. (\cite{RNW04}) \\			 
166473 & 7700& 4.20 & ~0.0   & 8.6 &Gelbmann et al. (\cite{GRW00})  \\	 
128898 & 7900& 4.20 & 12.5   & 1.5 &Kupka et al. \cite{KRW96}) \\	 
 29578 & 8000& 4.20 & ~2.5   & 5.6 &Ryabchikova et al. (\cite{RNW04}) \\  
116114 & 8000& 4.10 & ~2.5   & 6.2 &Ryabchikova et al. (\cite{RNW04}) \\  
137909 & 8000& 4.30 & ~2.5   & 5.4 &Ryabchikova et al. (\cite{RNW04}) \\  
 47103 & 8180& 3.50 & ~0.0   &16.3 &this paper \\			  
188041 & 8800& 4.00 & ~0.0   & 3.6 &Ryabchikova et al. (\cite{RLKB04}) \\ 
 66318 & 9200& 4.25 & ~0.0   &15.5 &Bagnulo et al. (\cite{bagn03}) \\	 
133792 & 9400& 3.70 & ~0.0   & 1.1 &Kochukhov et al. (\cite{KTRM06}) \\	 
118022 & 9500& 4.00 & 10.0   & 3.0 &this paper \\			 
170973 &10750& 3.50 & ~8.0   & 0.0 &Kato (\cite{Kato03}) \\		 
144897 &11250& 3.70 & ~3.0   & 8.8 &Ryabchikova et al. (\cite{RRKB06}) \\ 
\hline
\multicolumn{6}{c}{Reference stars} \\
\hline
 24711 & 7650& 4.00 & 18.5   & 0.0 & this paper  \\
 61421 & 6510& 3.96 & ~3.5   & 0.0 &Allende Prieto et al. \cite{Procyon}\\
\hline
\end{tabular}
\end{table*}

\section{Model atmosphere parameters}

Fundamental parameters of the program stars are given in Table~\ref{tbl1}. For most stars effective temperatures \teff\ and surface gravities \lgg\
were taken from the literature (last column of Table~\ref{tbl1}). For HD\,965, HD\,47103, HD\,118022, and HD\,134214 atmospheric
parameters were derived using Str\"omgren photometric indices (Hauck \& Mermilliod \cite{HM98})
with the calibrations by Moon \& Dworetsky
(\cite{MD85}) and by Napiwotzki et al. (\cite{N93}) implemented in the {\tt TEMPLOGG} code
(Rogers \cite{R95}). For HD\,75445, HD\,176232, and HD\,203932 effective temperatures were slightly corrected by fitting H$\alpha$
profile. The mean surface magnetic fields \bs\ were derived from the resolved and partially resolved Zeeman patterns. 
In all stars rotational velocities were estimated by fitting line profiles of the
magnetically insensitive Fe\i\ 5434.5 and 5576.1 \AA\ lines. Model atmospheres were calculated with the {\sc ATLAS9} code (Kurucz \cite{K93}).

\section{Ca stratification analysis}

Before performing careful study of the IR Ca\ii\ $\lambda$~8498 line profile, we have to investigate Ca abundance distribution
in Ap atmospheres. 
In all program stars Ca stratification was derived using a set of spectral lines in the optical region, for which no indication on
the significant isotopic shifts exists. Atomic parameters
of these lines, as well as the Ca\ii\ $\lambda$~8498 line, are given in Table\,\ref{strat-list}. Stratification analysis requires 
high accuracy not only for
the oscillator strengths but also for the damping parameters, because Ca has a tendency to be concentrated close to the 
photospheric layers where the electron density is high. In particular, it is important for Ca\ii\ lines. For 
Ca\ii\ $\lambda\lambda$ 3158, 3933, 8248, 8254, 8498 lines the Stark damping constants were taken from the paper by 
Dimitrijevi\'c \& Sahal-Br\'echot (\cite{Stark}), where semi-classical calculations as well as a compilation of the
experimental data were presented. 
For the rest of the Ca lines the Stark damping constants calculated by Kurucz (\cite{K93}) were used.          
The oscillator strengths were taken mostly from the laboratory experiments, and they are verified by the recent NLTE analysis of
calcium in late-type stars (Mashonkina et al. \cite{MKP07}). Because of the large range in effective temperatures and magnetic
field strengths, we could not use the same set of lines for all stars. 

\begin{table*}[!t]
\caption{A list of spectral lines used for the stratification calculations. The columns give the ion identification,
central wavelength, the excitation potential (in eV) of the lower level, oscillator strength ($\log\,{gf}$), 
the Stark damping constant, and the reference for the oscillator strength. \label{strat-list}}
\begin{center}
\begin{tabular}{lcrrrl}
\noalign{\smallskip}
\hline
\hline
Ion &Wavelength &\ei\,(eV)  &$\log\,{gf}$&$\log\,\gamma_{\rm St}$ & Ref.\\
\hline
Ca\ii&  3158.869&    3.123 &    0.241&  -4.90  &Teodosiou \cite{T}  \\ 
Ca\ii&  3933.655&    0.000 &    0.105&  -5.73  &Teodosiou \cite{T}  \\ 
Ca\i &  4226.728&    0.000 &    0.244&  -6.03  &Smith \& Gallagher \cite{SG} \\ 
Ca\ii&  5021.138&    7.515 &   -1.207&  -4.61  &Seaton et al. \cite{TB} \\ 
Ca\ii&  5339.188&    8.438 &   -0.079&  -3.70  &Seaton et al. \cite{TB}\\
Ca\i &  5857.451&    2.933 &    0.240&  -5.42  &Smith \cite{S}  \\ 
Ca\i &  5867.562&    2.933 &   -1.57~&  -4.70  &Smith \cite{S}  \\ 
Ca\i &  6122.217&    1.896 &   -0.316&  -5.32  &Smith \& O'Neil \cite{SN} \\ 
Ca\i &  6162.173&    1.899 &   -0.090&  -5.32  &Smith \& O'Neil \cite{SN} \\ 
Ca\i &  6163.755&    2.521 &   -1.286&  -5.00  &Smith \& Raggett \cite{SR} \\ 
Ca\i &  6166.439&    2.521 &   -1.142&  -5.00  &Smith \& Raggett \cite{SR} \\ 
Ca\i &  6169.042&    2.253 &   -0.797&  -5.00  &Smith \& Raggett \cite{SR} \\ 
Ca\i &  6169.563&    2.256 &   -0.478&  -4.99  &Smith \& Raggett \cite{SR} \\ 
Ca\i &  6449.808&    2.521 &   -0.502&  -6.07  &Smith \& Raggett \cite{SR} \\ 
Ca\i &  6455.598&    2.523 &   -1.340&  -6.07  &Smith \cite{S}  \\ 
Ca\ii&  6456.875&    8.438 &    0.410&  -3.70  &Seaton et al. \cite{TB} \\ 
Ca\i &  6462.567&    2.523 &	0.262&  -6.07  &Smith \& Raggett \cite{SR} \\ 
Ca\i &  6471.662&    2.526 &   -0.686&  -6.07  &Smith \& Raggett \cite{SR} \\ 
Ca\ii&  8248.796&    7.515 &    0.556&  -4.60  &Seaton et al. \cite{TB} \\ 
Ca\ii&  8254.721&    7.515 &   -0.398&  -4.60  &Seaton et al. \cite{TB} \\ 
Ca\ii&  8498.023&    1.692 &   -1.416&  -5.70  &Teodosiou \cite{T}  \\ 
\hline
\end{tabular}
\end{center}
\end{table*}

The Ca stratification analysis was performed using the step-function approximation of the abundance distribution
(for details see Ryabchikova et al. \cite{RLK05}). In a few cases the step-function approximation can not provide
an adequate description of the full set of spectral lines. The obvious reasons are the use of normal non-magnetic star 
atmosphere with homogeneous element distribution for a star with known abundance stratification, and a deviation of the
abundance distribution from the simple step-function. In cooler stars the range of  formation depth of the 
optical lines is different from the IR-triplet lines of interest, therefore Ca abundance in the upper atmospheric
layers derived from the optical lines may be not accurate enough for the description of  cores of IR lines. Also, continuum
normalization in the IR lines region is indirectly based on the adopted effective temperatures, which may introduce significant uncertainty and sometimes lead to a poor fit in the line wings. 

\begin{figure}[!th]
\centering
\figps{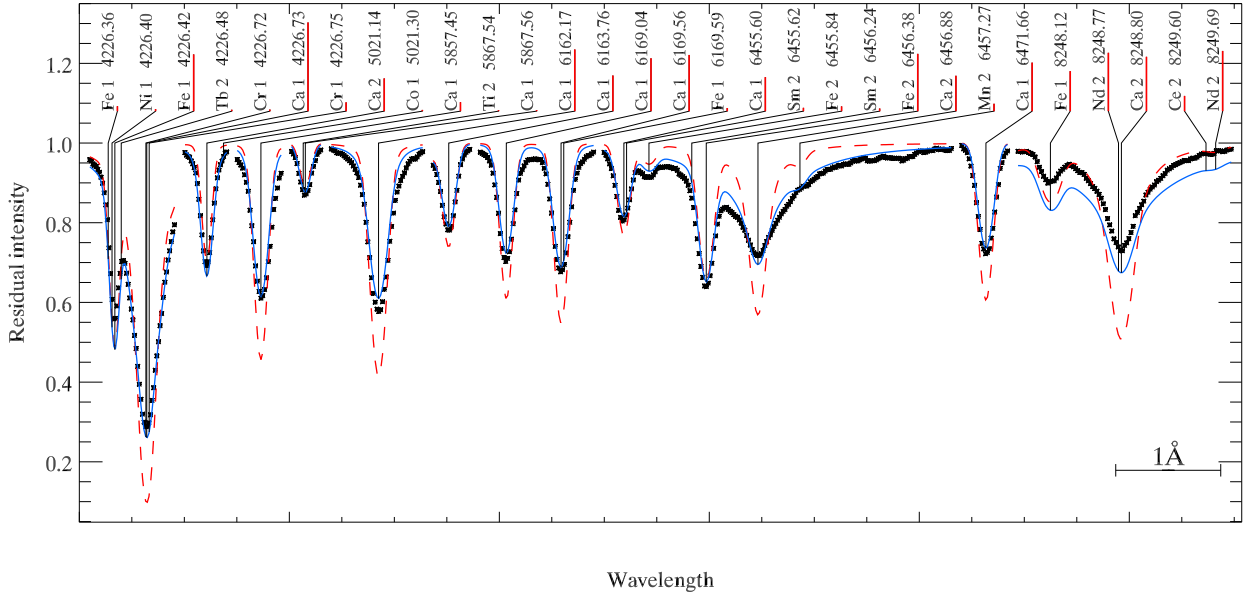}
\caption{A comparison between the observed line profiles (dots) and calculations with
the stratified Ca abundance distribution (full line)
and with the homogeneous Ca abundance (dashed line) in HD~176232. 
}
\label{10_Aql_Ca}
\end{figure}

We start the analysis
with the best homogeneous Ca abundance derived from a chosen set of spectral lines, and then vary parameters of the
step-function until the adequate fit to the observed line profiles is achieved. 
Magnetic spectral synthesis code {\tt SYNTHMAG}(Piskunov \cite{P99}; Kochukhov \cite{synthmag06}) was used in our calculations.
Fig.~\ref{10_Aql_Ca} displays the results of the stratification analysis for HD~176232 (10~Aql), where synthetic profiles calculated
with homogeneous Ca distribution $\log (Ca/N_{\rm tot})=-5.14$ are shown by dashed line while those calculated with
the stratified Ca distribution are shown by the full line. The derived Ca distribution is given in Fig.~\ref{HD176232_8498} (right panel).
The stratified Ca 
abundance yields two times smaller standard deviation compared to the homogeneous Ca distribution. 

The same procedure was applied to all stars
included our sample. Ca distributions in the atmospheres of several stars are shown in Fig.\,\ref{Ca_distr}. They are all characterized
by an abundance jump in the region $-1.3\ge\log\tau_{5000}\le-0.5$, an 1--1.5 dex overabundance deep in the atmosphere and 
a strong Ca depletion above  $log\tau_{5000}=-1.5$. It is difficult to say if there is any dependence on the effective temperature
and/or on the magnetic field strength.     

\section{Ca isotopic anomaly}

Ca has six stable isotopes, 40, 42, 43, 44, 46, 48, and in the solar-system matter Ca mixture consists mainly of $^{40}$Ca 
(96.9 \% - see Anders \& Grevesse \cite{AG89}). Table\,\ref{iso8498} gives wavelengths of all Ca isotopes following the
isotopic shifts measured by N\"ortersh\"auser et al. (\cite{Ca_IS}) as well as the isotopic fractional oscillator strengths
corresponding to the solar-system matter mixture.    

\begin{table*}
\caption{Atomic data for the isotopic components of Ca\ii\ 
$\lambda$~8498. The fractional isotope abundances $\epsilon$
correspond to the composition solar-system matter.}\label{iso8498}
\begin{center}
\begin{tabular}{ccc}
\hline \noalign{\smallskip}
 \ $\lambda,\AA$ & isotope  & $\log gf\epsilon$ \\
\noalign{\smallskip} \hline \noalign{\smallskip}
  8498.023 & 40 &  $-$1.43   \\
  8498.079 & 42 &  $-$3.60   \\
  8498.106 & 43 &  $-$4.29   \\
  8498.131 & 44 &  $-$3.10   \\
  8498.179 & 46 &  $-$5.81   \\
  8498.223 & 48 &  $-$4.14   \\
\noalign{\smallskip} \hline
\end{tabular}
\end{center}
\end{table*}

\begin{figure}[!t]
\centering
\fifps{80mm}{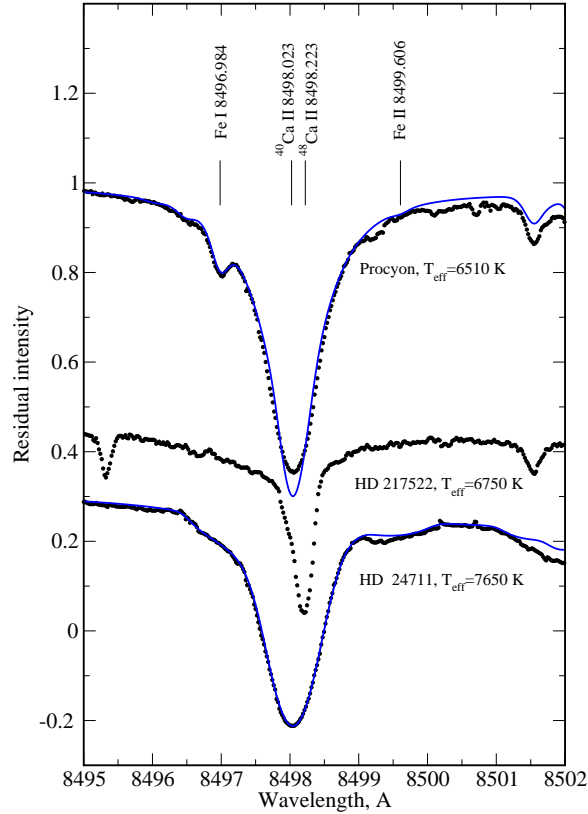}
\caption{A comparison between the observed line profiles (dots) and the calculations with
the solar-matter Ca isotopic mixture (full line) in the spectra of Procyon and Am star HD~27411.
The observed spectrum of Ap star HD~217522 is in the middle.}
\label{Ca_ref}
\end{figure}

With the solar-matter isotopic mixture we calculated Ca\ii\ $\lambda$~8498 line 
profile in the spectra of our reference stars Procyon and HD~27411 and compared them with the observations. Fig.~\ref{Ca_ref} shows
the results of this comparison. Although in the Procyon spectrum our LTE calculations cannot provide a very good fit, however, no
wavelength shift was detected in both stars. At the same time, the observed profile of this line in the spectrum of our program star
HD~217522 presented in Fig.\ref{Ca_ref} has a complex structure and is clearly redshifted with the strongest component being at the 
position of the heaviest Ca isotope.

\begin{figure}[!t]
\centering
\firrps{80mm}{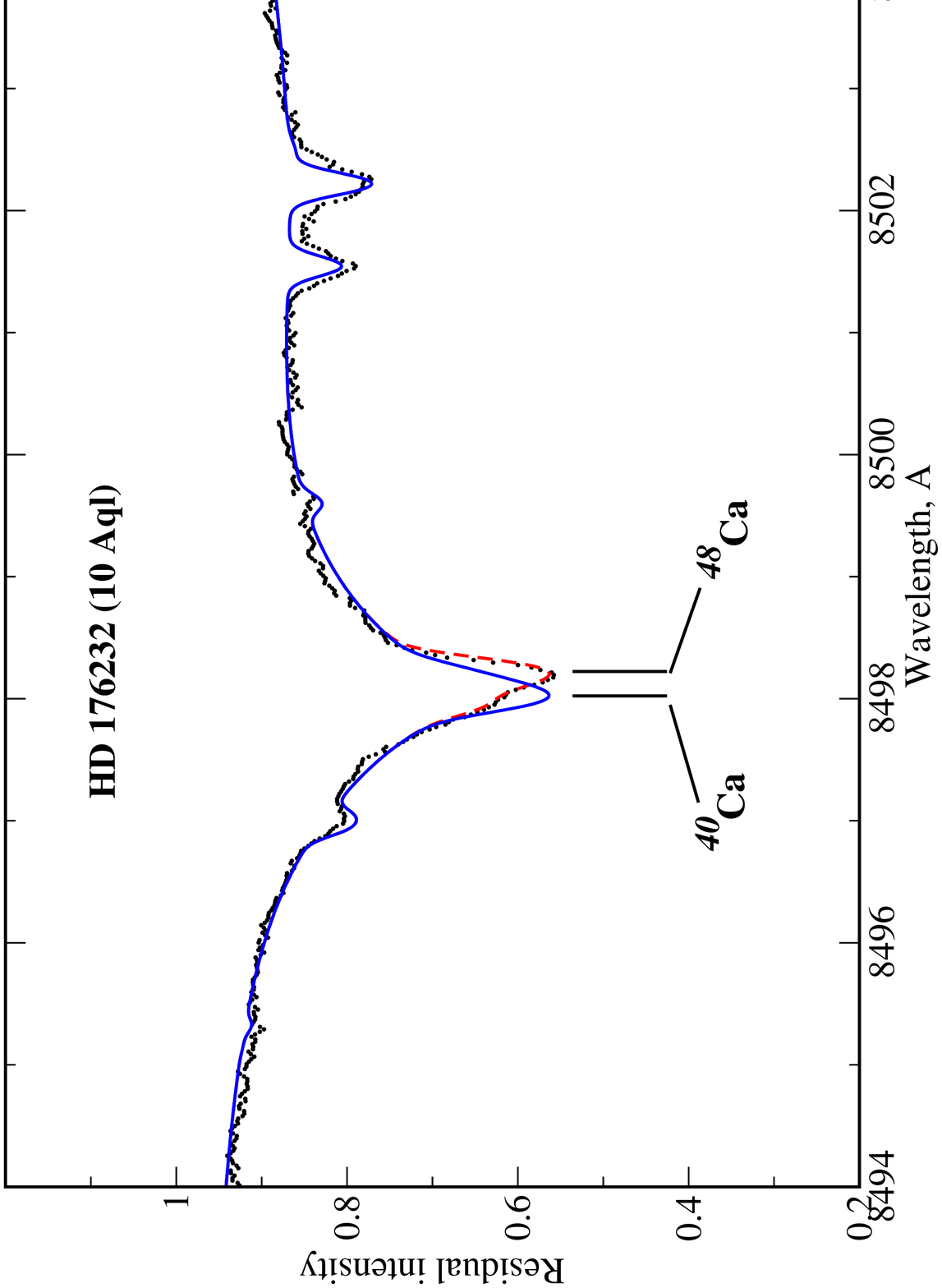}\hspace{0.5cm}\firrps{80mm}{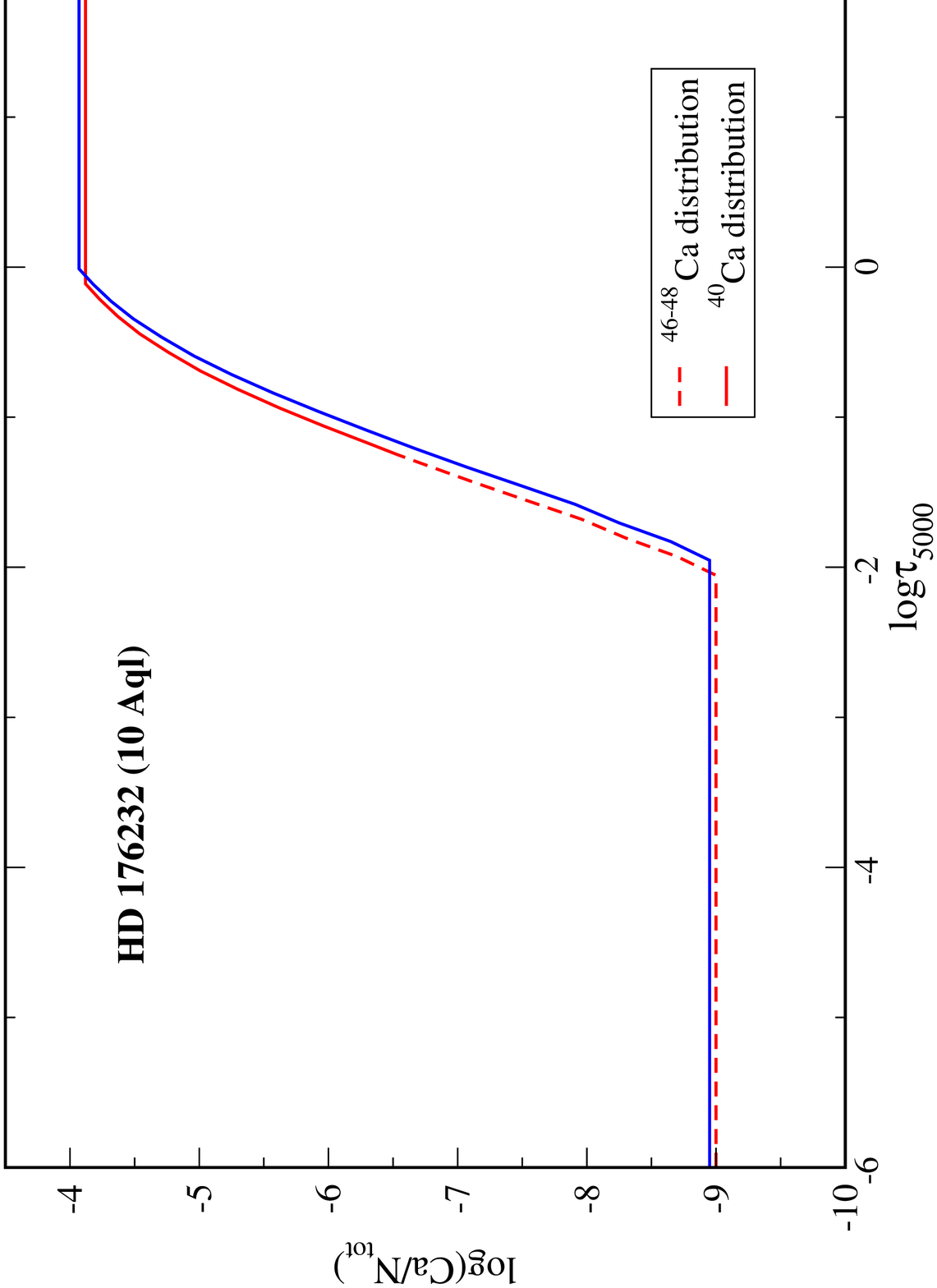}
\caption{A comparison between the observed (filled circles) and synthetic line profiles of Ca\ii\ $\lambda$~8498 (left panel), calculated
with Ca distribution shown in the right panel. Synthetic spectrum calculations with the solar-matter Ca isotopic mixture are shown by full
line, and those with Ca isotopic separation as indicated in the right panel are shown by dashed line. Ca distribution derived from the
optical lines is shown by the solid blue line in the right panel, while isotopic separation are shown by red line (dashed + solid). For
illustration purpose two distributions are slightly shifted relative to each other.}
\label{HD176232_8498}
\end{figure}

The core of the IR Ca\ii\ $\lambda$~8498 line is formed higher than any of the optical lines, except Ca\ii~$\lambda$~3933. For most
stars the Ca\ii\ $\lambda$~3933 line were not accounted in the stratification calculations, therefore Ca abundance in the upper atmosphere
may be rather uncertain, because all other optical lines are not sensitive to abundance variations above $log\tau_{5000}=-2.0$ to $-2.5$. The
abundance in the upper atmosphere is defined by the slope of the abundance gradient in the jump region. If the Ap atmosphere is close
to the normal {\sc ATLAS9} one (Kurucz \cite{K93}) adopted in our analysis, then  Ca\ii\ $\lambda$~8498 line should be fitted with 
the Ca abundance distribution derived from optical lines. Our calculations show that while it is correct for the observed total intensity,
in part of program stars  we cannot fit the line cores, which are often redshifted. Fig.~\ref{HD176232_8498} (left panel, dashed line)
shows a fit of synthetic spectrum calculated with the solar-matter Ca isotopic mixture and Ca abundance distribution (right panel) 
to the observed spectrum of HD~176232. One immediately notices that while the line wings are fitted rather satisfactory, the line core
cannot be fitted with the solar-matter Ca isotopic mixture. When we separate $^{40}$Ca and $^{46}$Ca, $^{48}$Ca isotopes in the atmosphere as indicated
in Fig.~\ref{HD176232_8498} (right panel), then we get a satisfactory agreement between the observed and calculated spectra (full line
in the left panel of Fig.~\ref{HD176232_8498}). Of course, it is a crude approximation, however it gives us a direct evidence of the 
Ca isotopic separation in the atmospheres of Ap stars. 

\begin{figure}[!t]
\centering
\fifps{100mm}{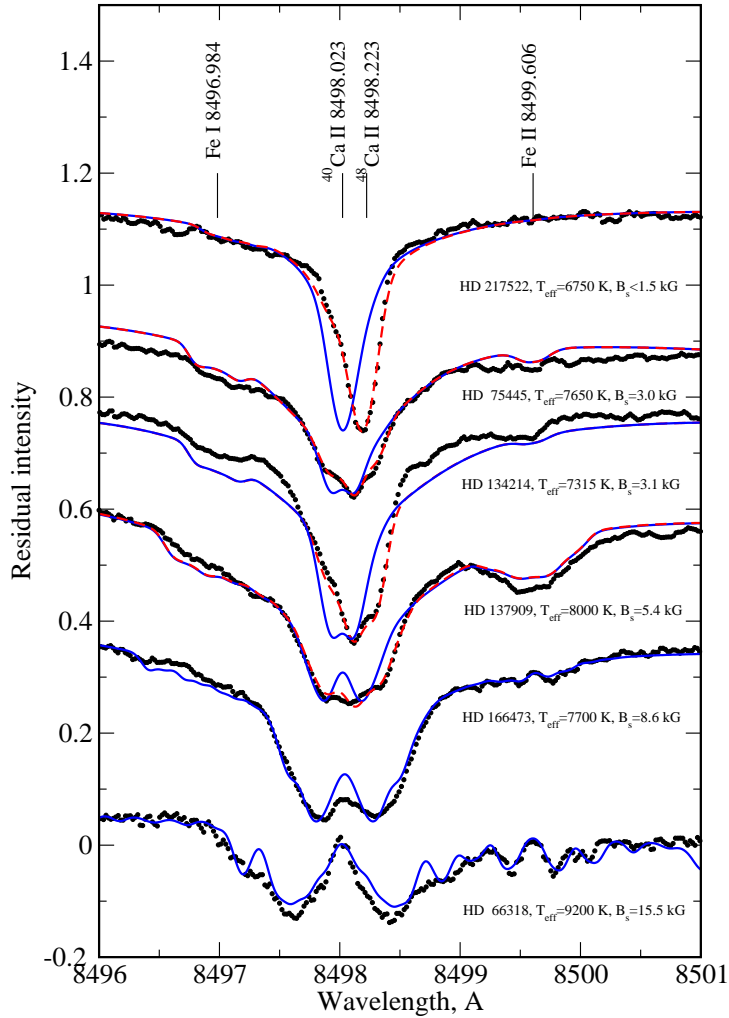}
\caption{The same as in Fig.~\ref{HD176232_8498} (left panel) but for a set of program stars.}
\label{8498_all}
\end{figure}

\begin{figure}[!t]
\centering
\firrps{100mm}{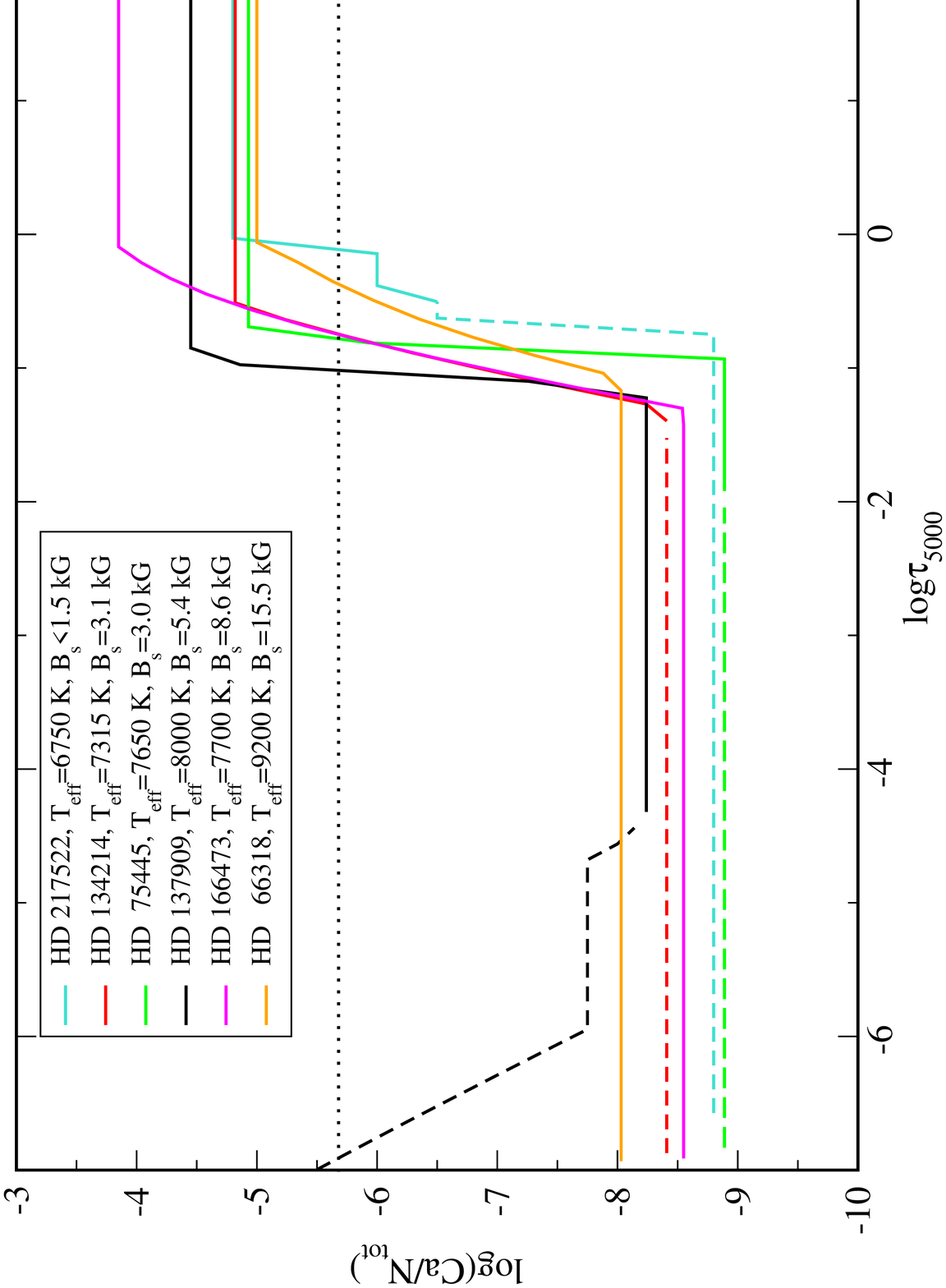}
\caption{Ca abundance distributions in the atmospheres of a few stars with different effective temperatures and 
magnetic field strengths. Dashed line shows optical regions where $^{40}$Ca is dominated, while the solid line indicates
the regions of predominant $^{46}$Ca and $^{48}$Ca isotopes concentration. Solar Ca abundance is marked by dotted line.}
\label{Ca_distr}
\end{figure}

This procedure was applied to all stars of our program. Fig.~\ref{8498_all} gives an example of our fitting procedure for a subset of
stars with different effective temperatures and different magnetic field strengths, and the corresponding Ca stratifications with the
isotopic separation are shown in Fig.~\ref{Ca_distr}.  In the stars with small to moderate magnetic fields we clearly see a significant
contribution of the heavy isotopes $^{46}$Ca and $^{48}$Ca, and this contribution decreases with the increase of the magnetic field strength.
Even in HD~137909 ($\beta$~CrB) with the mean magnetic modulus \bs=5.4 kG one still needs a small contribution of $^{48}$Ca, but under the
assumption of very specific Ca distribution shown in Fig.~\ref{Ca_distr}. 
We have to introduce a rapid increase of Ca abundance in a thin
upper atmospheric layer above $log\tau_{5000}=-5$. In principle, it does not contradict  the theoretical Ca  diffusion calculations.
Both Borsenberger at al. (\cite{BPM81} - Fig.~6) and Babel (\cite{Babel92}) obtained Ca abundance increase in the upper layers after
the abundance jump. However, NLTE treatment of the Ca lines formation is needed to investigate the upper atmospheric layers.  
 
The same results were obtained for other stars with similar magnetic field strengths: HD~965, HD~137949, HD~29578.

\section{Discussion}

If the overall distribution of Ca abundance in the atmospheres of Ap stars follows the predictions of the radiatively driven
diffusion, our results on the isotopic separation favour the light-induced drift (LID) as the main process responsible for 
this separation. Indeed, according to Atutov \& Shalagin (\cite{AS88}) LID arises when the radiation field is anisotropic inside the
line profile. Such an anisotropy takes place for a line of the trace isotopes, $^{46}$Ca, $^{48}$Ca for instance, in the solar-matter mixture,
which is sitting in the wing of a strong line of the main isotope $^{40}$Ca, and the main isotope should induce the drift velocity for
other isotopes. If we have  a trace isotope's line in the red wing of the main isotope's line, then the drift velocity is directed
from towards the upper atmosphere and the trace isotopes are pushed upwards. This is the case for the Ca isotopic structure. 
Zeeman splitting changes the line shape
and decreases the flux anisotropy for a trace isotope's line. When magnetic field becomes strong enough, $\sim4-5$ kG, then the flux
anisotropy disappears and the isotopic separation is ceasing. Therefore, the observed Ca isotopic anomaly in magnetic stars may be
qualitatively explained by the  combined action of the radiatively-driven diffusion and light-induced drift.

\begin{acknowledgements}
This work was supported by the RAS Presidium Program ``Origin and Evolution of Stars and Galaxies'', by Austrian Science 
Fund (FWF-P17580N2) and by grant 11630102 from the Royal Swedish Academy of Sciences. TR acknowledges the partial support from
RFBR grant 06-02-16110a and the Leading Scientific School grant 162.2003.02.
\end{acknowledgements}

\end{document}